The Fokker-Planck operator at a continuous phase transition

Moshe Schwartz

Raymond and Beverly Faculty of Exact Sciences
School of Physics and Astronomy
Tel Aviv University
Ramat Aviv 69978
Tel Aviv, Israel

I consider a physical system described by a continuous field theory and enclosed in a large but finite cubical box with periodic boundary conditions. The system is assumed to undergo a continuous phase transition at some critical point. The $\varphi^4$ theory that is a continuous version of the Ising model is such a system but there are many other examples corresponding to higher spin, higher symmetry etc. The eigenfunctions of the corresponding Fokker-Planck operator can be chosen, of course, to be eigenfunctions of the momentum operator. It is shown that the eigenvalues of the FP operator, corresponding to each eigenvalue **q** of the momentum operator, evaluated at a transition point of the finite system, accumulate at zero, when the size of the system tends to infinity. There are many reasonable ways of defining a critical temperature of a finite system, that tends to the critical temperature of the infinite system as the size of the system tends to infinity. The accumulation of eigenvalues is neither affected by the specific choice of critical temperature of the finite system nor by whether the system is below or above its upper critical dimension.

The property of critical slowing down [1,2] is known for a long time from experimental [3-7] and numerical work [8-11]. The quantitative description is in terms of characteristic decay times or alternatively "characteristic frequencies" $\omega_q$, that govern the decay of a disturbance of wave vector **q**. It was found that at the transition, $\omega_q$ behaves as some positive power of $q$, $\omega_q \propto q^z$. This implies that the larger the scale of the disturbance the longer it takes to decay and a divergent scale results in a divergent decay time. The evaluation of the exponent $z$ was the subject of theoretical work using a number of different approaches but all based on stochastic field equations of the Langevin type to describe the dynamics of the system [12-18]. More recently, it was suggested that not only do the characteristic frequencies tend to zero with $q$ but the decay function itself becomes slower than an exponential (e.g. stretched exponential) for long times, $\omega_q t >> 1$ [19-21]. That property was shown to hold for quite general Langevin field equations including in addition to critical dynamics, equations of the KPZ type. The derivation of stretched exponential decay or any other form of slow decay has to rely on approximations. As will become evident later (eq.(12)) the time dependent structure factor needs evaluations of the eigenstates and eigenvalues of a "quantum field Hamiltonian" (eq.(10)) obtained by a standard similarity transformation from the corresponding Fokker-Planck operator. The accuracy with which these have to be evaluated has to increase with the time argument of the structure factor. Therefore the problem of very long time decay is extremely difficult and it is important to have as many exact results as possible in support of approximate derivations. From eq. (12) it will become clear also that a necessary condition for slow decay of a disturbance of wave vector **q** is to have "enough" eigenvalues corresponding to states carrying momentum **q** in the vicinity of zero. It has been shown in a recent publication [20], that there must exist an eigenvector of the FP operator, at the transition point, that carries momentum, **q** and has an eigenvalue that is as close as we wish to zero. In the present article, I go beyond that and show that for each **q** the eigenvalues of the FP operator accumulate at zero for the infinite system. The proof I present relies on the simplest established lowest order results of finite size scaling equilibrium theory and does not involve even the next order corrections such as the dependence of a properly chosen critical temperature of the finite system on its size [22, 23]. To my knowledge, those results have never been proven rigorously. Nevertheless their use as conjectures in the present proof yields information about a non equilibrium problem that is considerably more difficult and that information is exact provided that those widely accepted equilibrium results are indeed correct.

Consider a system described in terms of a scalar field $\varphi$ and enclosed in a cube with periodic boundary conditions. The static statistical properties of the system, are assumed to be given by the Gibbs distribution, $P_{eq} \propto \exp[-W]$, where $W$ is the classical dimensionless Hamiltonian of the system. I have in mind Hamiltonians that are even functionals of the field and are generalizations of the $\varphi^4$ theory that in term of Fourier components of the field has the form

$$W = \frac{1}{2}\sum (r_0 + q^2)\varphi_{\mathbf{q}}\varphi_{-\mathbf{q}} + \frac{u}{\Omega}\sum \varphi_{\mathbf{l}_1}\varphi_{\mathbf{l}_2}\varphi_{\mathbf{l}_3}\varphi_{-(\mathbf{l}_1+\mathbf{l}_2+\mathbf{l}_3)}$$

(1)

where the Fourier components of the field defined by

$$\varphi_{\mathbf{q}} = \frac{1}{\sqrt{\Omega}}\int \varphi(\mathbf{r})\exp(-i\mathbf{q}\cdot\mathbf{r})\, d\mathbf{r}. \qquad (2)$$

and $\Omega$ is the volume of the system. The momentum indices are assumed to be bound from above by some high momentum cut off and their allowed values are given by

$$\mathbf{q} = \frac{2\pi\cdot\mathbf{n}}{L}, \qquad (3)$$

where $L$ is the linear size of the system and $\mathbf{n}$ a vector of integers. The following discussion and the proof I present is not limited, however, to the above typical model. The discussion clearly covers other systems with different dependences of $W$ on the field (e.g. systems describing higher spins) or systems in which the field is not a scalar (e.g. systems with O(n) symmetry). As will become evident, the only property that I use is that there is a critical point in the large volume limit and that at that point and in that limit the structure factor $<\varphi_{\mathbf{q}}\varphi_{-\mathbf{q}}>$ or its analog in the case of vector fields, diverges at small $q$'s as $q$ raised to a power smaller than -1.

The dynamics is assumed to be described in terms of a set of Langevin field equations

$$\frac{d\varphi_{\mathbf{q}}}{dt} = -\gamma \frac{\partial W}{\partial \varphi_{-\mathbf{q}}} + \eta_{\mathbf{q}}, \tag{4}$$

where the noise $\eta_{\mathbf{q}}$ obeys

$$\langle \eta_{\mathbf{q}}(t) \rangle = 0 \quad \text{and} \quad \langle \eta_{\mathbf{q}}(t)\eta_{-\mathbf{q}}(t') \rangle = 2\gamma \delta(t-t'). \tag{5}$$

A standard procedure leads to the Fokker-Planck equation for the probability distribution of the fields, $P$,

$$\frac{\partial P}{\partial t} = \gamma \sum_{\mathbf{q}} \frac{\partial}{\partial \varphi_{\mathbf{q}}} \left[ \frac{\partial}{\partial \varphi_{-\mathbf{q}}} + \frac{\partial W}{\partial \varphi_{-\mathbf{q}}} \right] P \equiv OP, \tag{6}$$

where $O$ is the FP linear operator. The distribution at equilibrium, $P_{eq}$ is the eigenstate of the Fokker-Planck operator, $O$, with eigenvalue zero. It follows directly from equation (6), that

$$P_{eq} \propto \exp[-W], \tag{7}$$

is the Gibbs distribution.

I will be interested in the time dependent structure factor $\langle \varphi_{-\mathbf{q}}(0)\varphi_{\mathbf{q}}(t) \rangle$, where the meaning of the average is as follows: $\varphi_{-\mathbf{q}}$ is measured at time $t=0$ at equilibrium, the system is then allowed to evolve freely and $\varphi_{\mathbf{q}}$ is measured at time $t$. The mathematical form of the above statement is

$$\langle \varphi_{-\mathbf{q}}(0)\varphi_{\mathbf{q}}(t) \rangle = \int D\varphi^1 D\varphi^2 \varphi^1_{-\mathbf{q}} P_{eq}\{\varphi^1\} \varphi^{(2)}_{\mathbf{q}} P\{\varphi^{(2)}; \varphi^{(1)}, t\}, \tag{8}$$

where $P_{eq}$ is normalized and $P\{\varphi^{(2)}; \varphi^{(1)}, t\}$ is the solution of the Fokker-Planck equation (5) for a distribution in the variables $\varphi^{(2)}$ with initial condition

$$P\{\varphi^{(2)};\varphi^{(1)},0\}=\prod_l \delta(\varphi_l^{(2)} - \varphi_l^{(1)}). \tag{9}$$

A standard transformation, $P = P_{eq}^{1/2}\psi$, which induces a similarity transformation on $O$, brings the Fokker-Planck eq. (6) and the definition (8) into forms more familiar from quantum mechanics.

$$\frac{\partial \psi}{\partial t} = \gamma \sum_q \left[ \frac{\partial}{\partial \varphi_q} - \frac{1}{2}\frac{\partial W}{\partial \varphi_q} \right]\left[ \frac{\partial}{\partial \varphi_{-q}} + \frac{1}{2}\frac{\partial W}{\partial \varphi_{-q}} \right]\psi \equiv -\mathcal{H}\psi, \tag{10}$$

and

$$\langle \varphi_{-q}(0)\varphi_q(t) \rangle = \langle 0 | \varphi_{-q} \exp[-\mathcal{H}t] \varphi_q | 0 \rangle. \tag{11}$$

The "Hamiltonian", $\mathcal{H}$, obtained by the similarity transformation from $O$, is Hermitian and all its eigenvalues are positive apart from the eigenvalue zero, which corresponds to the ground state $|0\rangle$. Since the system is translationally invariant, $\mathcal{H}$ commutes with the momentum operator so that the eigenstates of $\mathcal{H}$ can be chosen to carry definite momenta.

Let $\{|n_q\rangle\}$ be the set of eigenstates of $\mathcal{H}$ carrying momentum $\mathbf{q}$ and $\{\lambda n_q\}$ the corresponding eigenvalues. It is easily verified that

$$\langle \varphi_{-q}(0)\varphi_q(t) \rangle = \sum |\langle n_{-q} | \varphi_q | 0 \rangle|^2 \exp[-\lambda n_q t]. \tag{12}$$

(This follows from the fact that $\varphi_q|0\rangle$ is an eigenstate of the momentum operator with momentum $\mathbf{q}$). The spectrum of $\mathcal{H}$ limited to the states $\{|n_q\rangle\}$, $\{\lambda n_q\}$, has thus direct bearing on the decay of the time dependent structure factor. For example if the spectrum $\{\lambda n_{\mathbf{q}}\}$ has a gap $\Delta_{\mathbf{q}}$, the decay of the time dependent structure factor is faster than $e^{-\Delta_{\mathbf{q}} t}$.

Define next

$$\rho_{\mathbf{q}}(\lambda) = \sum |\langle n_{\mathbf{q}} | \varphi_{\mathbf{q}} | 0 \rangle|^2 \delta(\lambda n_{\mathbf{q}} - \lambda) / \langle \varphi_{-\mathbf{q}}(0) \varphi_{\mathbf{q}}(0) \rangle. \tag{13}$$

The time dependent correlation function can be written as

$$\langle \varphi_{-\mathbf{q}}(0) \varphi_{\mathbf{q}}(t) \rangle = \langle \varphi_{-\mathbf{q}}(0) \varphi_{\mathbf{q}}(0) \rangle \int_0^\infty d\lambda \rho_{\mathbf{q}}(\lambda) \exp[-\lambda t], \tag{14}$$

It is clear, that for a finite system, it will always be possible to find a region $\lambda > 0$ including the origin and depending on the size of the system, such that $\rho_{\mathbf{q}}(\lambda)$ vanishes identically in that region, implying a decay in time that is faster than some exponential. This may be characterized by a lifetime that depends on the size of the system. The question is how does $\rho_{\mathbf{q}}(\lambda)$ behave in the limit of an infinite system. If in that limit $\rho_{\mathbf{q}}(\lambda)$ does not vanish too fast as $\lambda$ tends to zero, the decay will be slower than exponential (e.g. stretched exponential, power law etc.) The definition of $\rho_{\mathbf{q}}(\lambda)$ implies that in order for it not to vanish too fast as $\lambda$ tends to zero two conditions have to be met. First, the eigenvalues $\lambda n_{\mathbf{q}}$ have to accumulate at zero as the size of the system tends to infinity. The second condition is that the matrix elements in eq.(13) do not vanish too fast as the corresponding eigenvalues tend to zero. The actual long time dependence of the decay is strongly affected by the behavior of those matrix elements. The accumulation of eigenvalues to be discussed next is thus a necessary condition for slow decay but it does not determine the form of the decay. In fact, it is not even sufficient. The structure factor in the linear system to be discussed in the following as a prelude to the general case decays exponentially in spite of the relevant eigenvalues accumulating at zero. The reason is that $|\mathbf{n}^{(1)}_{\mathbf{q}}\rangle = \varphi_{\mathbf{q}} | 0 \rangle / \langle \varphi_{\mathbf{q}} \varphi_{-\mathbf{q}} \rangle^{1/2}$ is an exact normalized eigenstate of the Hamiltonian so that apart from the matrix element $\langle \mathbf{n}^{(1)}_{\mathbf{q}} | \varphi_{\mathbf{q}} | 0 \rangle$ all other matrix elements in eq. (12) vanish. The proof I'll present concerns only the first necessary condition that as

the size of the system tends to infinity the eigenvalues of *O* accumulate at zero for each **q**. This has no bearing on the actual form of long time decay.

The strategy of the proof is as follows: (a) The set of many "excitation" states carrying momentum **q** is defined and the corresponding expectation values of the "Hamiltonian" $\mathcal{H}$ are obtained and shown to accumulate at zero for any **q**. (b) The relation between those expectation values and the expectation values in an orthogonal set of states corresponding to the original "almost orthogonal" set of many "excitation" states is outlined. (c) It is shown how the above implies that the eigenvalues of the "Hamiltonian" carrying momentum **q** accumulate at zero.

Before proceeding to the proof in the general case, I'll describe first a number of common results from equilibrium theory that will be used as conjectures in the proof and define some useful definitions. Then I'll show how the accumulation of eigenvalues of $\mathcal{H}$ works for the soluble, linear case. This will give some hints and motivations for the general non-linear case.

(1) I keep the value of **q** fixed (eq. 3) as the size of the system is increased by considering only multiplication of the original size, *L*, by an integer. (2)There are many reasonable ways to define the critical temperature of the finite system. It may be taken to be the temperature at which the equilibrium structure factor at the smallest possible nonzero **q** is maximal. It may be chosen as the temperature where the specific heat or one of its derivatives with respect to temperature is maximal. For our purpose it is enough even to choose it as the transition temperature of the infinite system. (4) Regardless of the choice of the finite size critical point the structure factor has the scaling form.

$$\langle \varphi_\mathbf{q} \varphi_{-\mathbf{q}} \rangle = A q^{-(2-\eta)} \quad \text{for } qL \gg 1, \tag{15}$$

to leading order in *L* and for *q* small compared to the cut-off momentum $q_0$. Note that since *q* is kept fixed the condition $qL \gg 1$ can always be attained. Note that that is true also above the upper critical dimension where $\eta=0$ regardless of the fact that hyper scaling does not hold. Higher order correlation functions will also be needed. I will need $\langle \varphi_{\mathbf{k}_1} ... \varphi_{\mathbf{k}_m} \rangle$, where *m* is even and $\sum \mathbf{k}_i = 0$ but no subset of the **k**'s sums

up to zero. For a large finite system, these correlations have at the critical point the scaling form

$$\left|\langle \varphi_{\mathbf{k}_1}...\varphi_{\mathbf{k}_m}\rangle\right| = A(m)\frac{1}{L^{d(m-2)/2}} f(\mathbf{k}_1,...,\mathbf{k}_m), \tag{16}$$

to leading order in $L$. For $|\mathbf{k}|$'s small compared to $q_0$, that obey also $|\mathbf{k}|L \gg 1$ the function $f$ is homogeneous. Namely, if we scale all its variables by $\alpha$, the function of the scaled variables is related to the function of the original variables by

$$f(\alpha\mathbf{k}_1,...,\alpha\mathbf{k_m}) = \alpha^{-m(1-\eta/2)} f(\mathbf{k}_1,...,\mathbf{k}_m). \tag{17}$$

The pre factor $A(m)$ is of combinatorial origin and we will need to assume that there exists a finite $b$ such that $\ln A(m) < (bm)\ln(bm)$.

(4) A set of momenta, $S_k = \{\mathbf{l}_1,...,\mathbf{l}_k\}$, that can contain, in principle, the same momenta a number of times, is said to be irreducible if it does not contain real subsets, the momenta of which sum up to zero.

(5) A set $\overline{S_k}$ conjugate to $S_k$ is the set $\{-\mathbf{l}_1,...,-\mathbf{l}_k\}$.

(6) The sum $S_1 \oplus S_2$ of the sets $S_1$ and $S_2$, contains all momenta appearing in $S_1$ and $S_2$. The number of times each momentum appears in the sum is the sum of the number of times it appears in each set separately.

(7) The many "excitation" set is the set of states of the form

$$\psi_q(\mathbf{l}_1,...,\mathbf{l}_m) = \prod_{i=1}^{m} \varphi_{\mathbf{l}_i}|0\rangle \equiv \Psi_\mathbf{q}|0\rangle, \tag{18}$$

where $\sum_{i=1}^{m}\mathbf{l}_i = \mathbf{q}$ and where the set $\{\mathbf{l}_1,...,\mathbf{l_m}\}$

is irreducible.

A "physical proof" would probably proceed along the following lines. First the $\mathbf{l}_i$"s are chosen to equal $\mathbf{q}/m$. Then eqs.(21) and (22) are used to obtain the corresponding expectation value, $q^{(2-\eta)}/(Am^{(1-\eta)})$. The third step would be to assert that because of eq.(17), the "many excitation" states form an orthogonal set in the large volume limit

(that is also orthogonal to the ground state at any finite size , because it carries a different momentum), the eigenvalues of the "Hamiltonian" $\{\lambda n_q\}$ accumulate at zero. The "mathematical proof" proceeds basically along the same lines but has to take into account two difficulties. The first is that if **q** is an allowed momentum, **q**/*m* is generically not allowed. The second is much more serious. The assertion that the set of representative states is orthogonal in the large volume limit is problematic. Many examples exist in which the fact that the scalar products of normalized states tend to zero with the size of the system does not imply consequences expected from orthogonal sets in finite systems.

The proof for the Gaussian case (*u*=0) is straight forward. In that case, the states (18) are exact eigenstates of the "Hamiltonian" $\mathcal{H}$ with eigenvalues.

$$\lambda_q^m(\mathbf{l}_1,...,\mathbf{l}_m) = (\gamma/2)\sum l_i^2 , \qquad (19)$$

In the following I will introduce certain restrictions on the size of *m* , relating it to *q* . I will use these restrictions for the Gaussian case as well for the general one. These do not limit the final result about the accumulation of the eigenvalues but enable an exact proof under the conjectures about equilibrium correlation functions at the transition.

For any given **q** the size of the system can be chosen large enough so that *n*, the absolute value of the vector **n** corresponding to **q**, is large enough so that after choosing a number $0 < \theta < \min(1/2,\theta_0)$, where $\theta_0$ will be specified later (eq. (25)), the integers *m* in the range $(n^\theta, 1.5n^\theta)$ are large enough , so that the following will hold: (a) It is possible to choose for each *m* a representative state of the form (18) such that the sizes of the vectors $\mathbf{l}_i^m$ defining that representative are bound from above by *q*/(*m*-1).(The idea is to choose all the $\mathbf{l}_i$'s as close as possible to **q**/*m* since the choice $\mathbf{l}_i = \mathbf{q}/m$ for all *i* yields the lowest possible value for the eigenvalue in the sector of "*m* excitations" states. The above obvious choice is impossible, however, because if **q** is an allowed momentum **q**/*m* is generically not allowed. The conditions on *m* ensure that it is possible to choose the $\mathbf{l}_i$'s in such a way that the corresponding eigenvalue (eq.(19)) departs in a controlled way from the absolute minimum of the expression ).b) Furthermore, the set $Sm_1 \oplus Sm_2$ is irreducible for any

$m_1$ and $m_2$ in the range. (Note that this is not required for the proof in the Gaussian case but will simplify the proof in the general case.) Consequently, the representative eigenvalues are bound from above

$$\lambda^0_{q,rep}(\mathbf{l}_1,...,\mathbf{l}_m) \leq (\gamma/2)\frac{m}{(m-1)^2}q^2 \leq (\gamma/2)\frac{q^2}{n^\theta}, \quad \text{to leading order in } n. \tag{20}$$

Since $n = qL/2\pi$, the meaning of eq. (20) is that there are at least $\frac{1}{2}(qL/2\pi)^\theta$ states with eigenvalues smaller than $(\gamma/2)\frac{q^2}{(qL/2\pi)^\theta}$. Namely, as the size of the system tends to infinity, the eigenvalues of $\mathcal{H}$ accumulate at zero for any $\mathbf{q}$.

The proof for the general case will proceed in two steps. The first is to show that the expectation values of $\mathcal{H}$ in the many "excitations" states (18) accumulate at zero. The second is to show that this accumulation implies that the eigenvalues of $\mathcal{H}$ also accumulates, at zero for any $\mathbf{q}$ when the size of the system tends to infinity.

The representative states are exactly those chosen before for the Gaussian case. The corresponding expectation values of the $\mathcal{H}$ are

$$\mu_{q,rep}(\mathbf{l}_1,...,\mathbf{l}_m) = \frac{1}{2}\langle 0|[\prod_{i=1}^m \varphi_{-\mathbf{l}_i},[\mathcal{H},\prod_{i=1}^m \varphi_{\mathbf{l}_i}]]|0\rangle / \langle 0|\prod_{i=1}^m \varphi_{-\mathbf{l}_i}\varphi_{\mathbf{l}_i}|0\rangle$$

$$= \gamma \sum_{j=1}^m \langle 0|\prod_{i \neq j}^m \varphi_{-\mathbf{l}_i}\varphi_{\mathbf{l}_i}|0\rangle / \langle 0|\prod_{i=1}^m \varphi_{-\mathbf{l}_i}\varphi_{\mathbf{l}_i}|0\rangle. \tag{21}$$

The above is an exact expression that holds for a system of any size. The main points that lead to the first equality above are invariance under reflection ($\langle 0|\Psi_{-\mathbf{q}}\mathcal{H}\Psi_{\mathbf{q}}|0\rangle = \langle 0|\Psi_{\mathbf{q}}\mathcal{H}\Psi_{-\mathbf{q}}|0\rangle$ and $\mathcal{H}|0\rangle=0$. The second equality in (21) above is due to the fact that the double commutator $[\Psi_{-\mathbf{q}},[\sum \frac{\partial^2}{\partial\varphi_\mathbf{l}\partial\varphi_{-\mathbf{l}}},\Psi_\mathbf{q}]]$ equals $2\sum \frac{\partial\Psi_\mathbf{q}}{\partial\varphi_\mathbf{l}}\frac{\partial\Psi_{-\mathbf{q}}}{\partial\varphi_{-\mathbf{l}}}$. It may seem that the last expression on the left hand side of (21) does not depend on the non-linear coupling $u$ in the "Hamiltonian". This is not true of course, because the ground state $|0\rangle$ depends on $u$. To leading order in the size of the system

$$\langle 0|\prod_{i=1}^{m}\varphi_{-\mathbf{l}_i}\varphi_{\mathbf{l}_i}|0\rangle = \langle 0|\varphi_{\mathbf{l}_j}\varphi_{-\mathbf{l}_j}|0\rangle\langle 0|\prod_{i\neq j}^{m}\varphi_{-\mathbf{l}_i}\varphi_{\mathbf{l}_i}|0\rangle, \tag{22}$$

The corrections are terms of the same order of magnitude multiplied by a combinatorial factor of the order of $m^2$ and divided by the volume of the system, the correction is of the relative order of $L^{2\theta-d}$, that tends to zero for any d (the dimensionality of the system) as the size of the system tends to infinity. Thus, for a large enough system, the approximation (22) can be used to yield

$$\mu_{q,rep}(\mathbf{l}_1,...,\mathbf{l}_m) \leq \frac{\gamma}{A}[(3/2)n]^{\theta}\left(\frac{q}{n^{\theta}}\right)^{2-\eta}. \tag{23}$$

(Note that $qL$ here is still large in spite of the large $m$. In fact it is $2\pi n^{(1-\theta)}$ and therefore much larger than 1 so that eq.(15) is used for the structure factor.) This means, that there are $\frac{1}{2}(qL/2\pi)^{\theta}$ states, at least, for which the expectation values of $\mathcal{H}$ are smaller than $\frac{\gamma}{A}(3/2)^{\theta}\frac{q^{2-\eta}}{(qL/2\pi)^{(1-\eta)\theta}} \equiv \varepsilon(q,L)$. Since $\eta$ is always smaller than 1 this implies that as the size of the system tends to infinity, the expectation values of $\mathcal{H}$ accumulate at zero for any given $\mathbf{q}$. Note that so far I have used only the scaling form of the two point function (eq.(15)) as the only input conjecture in the proof.

If the representative states above were orthogonal, it would be easy to complete the proof that the eigenvalues of $\mathcal{H}$ corresponding to states carrying momentum $\mathbf{q}$ accumulate at zero. This, however, is not the case. The purpose of the next part of the proof is to show that the set of representative states, described above, can be used to construct an orthogonal set of states, by a Grahm-Schmidt orthoganolization procedure, such that the expectation values of $\mathcal{H}$ in the states of that set are identical to leading order in the size of the system to the corresponding expectation values calculated above. Because of the irreducibility of any set, $Sm_1 \oplus \overline{Sm_2}$ the scalar products of pairs of the representative states are given by eq. (16). Denote, for simplicity the many "excitation" representative states by $\psi_m$ and the smallest and largest $m$'s between $n^\theta$ and $1.5n^\theta$ by $m_s$ and $m_l$ respectively. Further denote by max and min the maximal and minimal values respectively of $|(\psi_i, \psi_j)|$ for $i \neq j$ in the range $m_s \leq i, j \leq m_l$.

The orthogonal set $\{\chi_m\}$ is obtained by constructing linear combinations of the $\psi's$

$$\chi_i = \psi_i + \sum_{j=m_s}^{i-1} a_{ij} \psi_j, \qquad (24)$$

It is not difficult to show that if min can be made as large as we wish compared to $[\max]^2 n^\theta / A^{n^\theta} L^{(2-\eta)n^\theta}$ when the size of the system is increased, $a_{ji}$ is given to leading order in the size of the system by $-\dfrac{(\psi_i, \psi_j)}{(\psi_i, \psi_i)}$ for $j>i$. Use of equations (15) and (16) enables to show that the inequality relating min and max can be made to be obeyed for large enough systems. It is easy to show now that the required inequality relating max and min is obeyed if

$$\theta < (d - 2 + \eta)/2b \equiv \theta_0. \qquad (25)$$

This choice of the range of $\theta$ ensures that for $i \neq j$ in the range, $|(\psi_i, \psi_j)|$ is monotonically decreasing as a function of $i + j$, for large enough $L$. It also ensures that,

$$(\chi_i, \chi_i) = (\psi_i, \psi_i) \quad \text{and} \quad (\chi_i | \mathcal{H} \chi_i) = (\psi_i, \mathcal{H} \psi_i), \qquad (26)$$

where relative corrections are at most of the order of $[\max]^2 n^\theta / [\min^2 A^{n^\theta} L^{(2-\eta)n^\theta}]$

(In fact, it may seem that a weaker condition on $\theta$, $\theta < (d-1+\eta/2)/2b$ ($\eta$ is always smaller than 2), suffices, to obtain eq. (26) but since the monotonicity of $|(\psi_i, \psi_j)|$ as a function of $(i+j)$ simplifies matters considerably, the somewhat more stringent limitation is used.) This implies that to leading order in $L$, the expectation values of $\mathcal{H}$ in the orthogonal set $\{\chi_m\}$ are the same as the expectation values in the many "excitation" states. Note that in the proof of the existence of the orthogonal set that is "close" to the non orthogonal set of many excitations I used the input scaling form of the many point correlation functions (eq.(17)) . The final stage of the proof is to show that the accumulation of the eigenvalues of the "Hamiltonian" follows. Assume that this is not the case. If so since the eigenvalues of $\mathcal{H}$ are positive, there exists a finite gap, $\Delta_\mathbf{q}$, independent on the size of the system (for large enough $L$) and there are only a finite number $M$ (independent on the size of the system) of eigenstates of $\mathcal{H}$ with energies in the gap that tend to zero with the size of the system. Denote these normalized eigenstates by $\Phi_i$ and the corresponding eigenvalue b$\lambda_q^i$ and any normalized vector vector belonging to the subspace spanned by eigenstates of the "Hamiltonian" with energies larger than the gap by $\Phi_\Delta$. Consider next the states $\{\chi_i\}$ with expectation values of the Hamiltonian smaller than $\varepsilon(q,L)$. I will express each of these states by

$$\chi_i = \sum b_{ij} \Phi_j + b_i \Phi_\Delta^i . \tag{27}$$

The expectation value $\mu_i$ corresponding to $\chi_i$ is bounded by

$$\mu_i \geq \sum |b_{ij}|^2 \lambda_q^j + |b_i|^2 \Delta_q \tag{28}$$

On the other hand $\mu_i < \varepsilon(q,L)$. The conclusion is that $|b_i|^2 < \dfrac{\varepsilon(q,L)}{\Delta_q}$. Therefore, the projections of the states $\chi_i$ on the $M$ dimensional space spanned by the $\Phi_i$'s are almost normalized and almost orthogonal to each other. The last result is impossible of course in view of the enormous number of sates $\chi_i$ with expectation values less than $\varepsilon(q,L)$. This completes the proof.

The proof given here concerns only the accumulation of eigenvalues of the FP operator at zero in the thermodynamic limit. This result is exact and general under the conjecture that the established results for equilibrium correlation functions at the transition are correct. It is thus a proof of the existence of a necessary condition for slower than exponential decay at the transition point. It has no bearing however on the actual form of decay whether exponential, stretched exponential or power law. The form of the decay depends on certain matrix elements that are problem specific. I hope to discuss this in future work. An interesting question that has been raised a number of times in connection with the above description is whether it is relevant to the glass transition. The first tendency is to say that it is irrelevant, because in the glassy state translational symmetry is broken and that was used heavily in the above. Furthermore, the glassy phase is supposed not to be in equilibrium and the above treatment is based on equilibrium dynamics. It seems, however, that there is some chance of success by considering the liquid phase close to the transition. I hope to come back to this in the near future.